\documentclass[letterpaper,10pt]{revtex4}
\usepackage{epsfig}

\newcommand{\be}{\begin{equation}}
\newcommand{\ee}{\end{equation}}
\newcommand{\ba}{\begin{eqnarray}}
\newcommand{\ea}{\end{eqnarray}}
\def\lsim{\mathrel{\rlap{\lower4pt\hbox{\hskip0pt$\sim$}}
    \raise1pt\hbox{$<$}}}
\def\gsim{\mathrel{\rlap{\lower4pt\hbox{\hskip0pt$\sim$}}
    \raise1pt\hbox{$>$}}}

\begin{document}

\title{Robustness of bipartite Gaussian entangled beams propagating in lossy channels}
\vskip 0.2in
\author{F. A. S. Barbosa$^1$, A. S. Coelho$^1$, A. J. de Faria$^1$, K. N. Cassemiro$^2$, A. S. Villar$^{2,3}$,
P. Nussenzveig$^1$, and M. Martinelli$^1$}
\vskip 0.1in
\address{$^1$ Instituto de F\'\i sica, Universidade de S\~ao Paulo,
Caixa Postal 66318, 05315-970 S\~ao Paulo, SP, Brazil. \\
$^2$ Max Planck Institute for the Science of Light, G\"unther-Scharowsky-str. 1 / Bau 24, 91058 Erlangen, Germany. \\
$^3$ University of Erlangen-Nuremberg, Staudtstr. 7/B2, 91058 Erlangen, Germany.} 
\vskip .6pc
\email{mmartine@if.usp.br}
\vskip 0.2in
\noindent

\maketitle

{\bf
Subtle quantum properties offer exciting new prospects in optical communications. Quantum
entanglement enables the secure exchange of cryptographic keys~\cite{QKD} and the distribution of
quantum information by teleportation~\cite{KimbleTele,KimbleQuInternet}. Entangled bright beams of
light attract increasing interest for such tasks, since they enable the employment of well-established
classical communications techniques~\cite{PKLamNPhoton}. However, quantum resources are fragile
and undergo decoherence by interaction with the environment. The unavoidable losses in the communication
channel can lead to a complete destruction of useful quantum properties -- the so-called ``entanglement
sudden death''~\cite{Eberly,UFRJ,Science09}. We investigate the precise conditions under which this
phenomenon takes place for the simplest case of two light beams and demonstrate how to produce states
which are robust against losses. Our study sheds new light on the intriguing properties of quantum
entanglement and how they may be tamed for future applications.
}

\vskip .6pc

Quantum entanglement is a counter-intuitive feature first introduced by Einstein, Podolsky, and
Rosen (EPR)~\cite{epr35} and discussed by Schr\"odinger~\cite{schrodPCS35} back in 1935. Beyond
its philosophical implications and fundamental character, it has increasing importance in proposals to
boost the processing power of computers and to make communications more secure. Bright beams
of light can be described in terms of physical observables -- the amplitude and phase quadratures --
analogous to the position and momentum of a particle as in the
original EPR conundrum. These continuous variables may be entangled and then
used for quantum key distribution or for quantum teleportation. Among all quantum states,
an important class is the one presenting Gaussian statistics, which have been extensively
investigated both theoretically and experimentally.

In the realm of quantum optics, squeezed states of light are an excellent example of non-classical
Gaussian states. They have quadrature fluctuations smaller than the classical limit of a coherent state.
It is well known that squeezing is degraded under channel losses,
an unwelcome effect for communications. Squeezed states always remain squeezed for
partial losses, linearly approaching the classical limit for complete attenuation~\cite{bachor}.
Only recently the effect of channel losses was analyzed for entanglement in continuous variables~\cite{Science09}.
As in the discrete scenario, entanglement can behave differently from the properties of each individual
system: it can vanish completely even for partial losses, a situation very similar to entanglement
sudden death (ESD) in two-qubit systems~\cite{Eberly,UFRJ,Science09}. In order to understand and devise
ways of controlling this effect in practical applications, we investigate the simplest and most fundamental situation
of two entangled Gaussian beams. We pinpoint the conditions leading to bipartite entanglement sudden death and trace a
boundary between states robust against channel losses and those subject to ESD.

We begin by describing our source of entangled light, the optical parametric oscillator (OPO)
operating above threshold. In this system, the nonlinear process of parametric downconversion
is stimulated, generating gain of the twin beams. As gain overcomes losses,
the system oscillates and outputs bright twin beams of light, with classical coherence
resembling that of a laser. The twin beams are entangled in their quadrature amplitude and phase
components~\cite{ReidDrumm88,prl05}. This can be understood from the energy conservation:
on the one hand, photons are created in pairs, implying strong intensity correlations;
on the other hand, the sum of their optical frequencies (thus their phase fluctuations)
has to equal the pump frequency, leading to phase anti-correlations. However, this simple picture is somewhat upset
by the existence of phonon noise in the crystal, which degrades the phase quantum
correlations and therefore hinders entanglement~\cite{pra09}. Control over the effects of
this noise enables the investigation of quantum state robustness.

\begin{figure}[htb]
   \centering
   \includegraphics[width=4in]{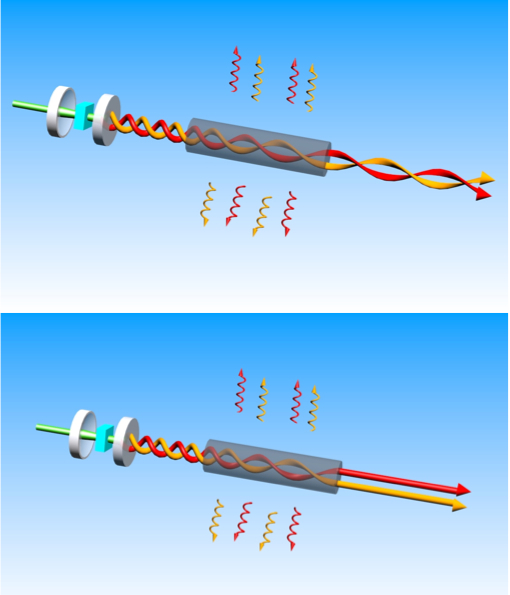}
   \caption{{\bf Pictorial view of the process:} An entangled state of two light beams (shown
   intertwined) in transmission through a lossy channel. At the output, the state may remain entangled
   (above) or become disentangled (below). }
   \label{fig:ESDart}
\end{figure}

To address the issue of sudden death, we need a necessary and sufficient entanglement criterion.
For bipartite Gaussian states, the positivity under partial transposition (PPT) fulfills this
requirement~\cite{simon}. Quantum properties of Gaussian states are completely characterized
by their second-order moments (variances and covariances), which can be conveniently organized in
the form of a covariance matrix. The PPT criterion can be stated in terms of the smallest symplectic eigenvalue
of the covariance matrix corresponding to the partially transposed state: if smaller than one, the matrix
represents an entangled state; otherwise, a separable state~\cite{simon}.
In order to apply the PPT test, one needs to completely reconstruct the covariance matrix.
Our study is thus conceptually simple: we generate twin light beams from the OPO, perform
a complete set of quadrature measurements and test the covariance matrix obtained for
entanglement. We then subject one of the beams to a controlled attenuation, simulating
propagation losses in a quantum channel, and repeat the procedure.
For each attenuation the symplectic eigenvalue reveals whether the beams remain entangled.
The situation is illustrated in Fig.~\ref{fig:ESDart}.

The states produced by the OPO are a physical realization of an EPR-type state,
i.e. the intensity difference ($\hat{p}_-$) and the phase sum ($\hat{q}_+$) of the
light beams show squeezing. Such states violate a simple inequality, derived
by Duan {\em et al.}~\cite{duan},
\begin{equation}
\Delta^2\hat{p}_- + \Delta^2\hat{q}_+ \ge 2 \; ,
\label{eq:DuanManco}
\end{equation}
\noindent
where the quadrature variance of a coherent state, the standard quantum limit (SQL), is unity.
Violation of this inequality is sufficient for entanglement. We will refer to it henceforth
as the ``Duan inequality''. It is very convenient to check experimentally, for it only
requires measurements of joint amplitude and phase correlations of both fields.
However, it is not a necessary criterion in this form, i.e. fulfillment of Eq.~(\ref{eq:DuanManco})
does not imply separability. In its sufficient and necessary form, for which
the complete covariance matrix has to be determined, the Duan criterion and PPT are equivalent~\cite{duan}.
We employ the less general Duan inequality of Eq.~(\ref{eq:DuanManco}), however, for its appealing
connection with the robustness of bipartite entangled states.

It is straightforward to check that two-mode entangled states and, more generally, states which violate
the Duan inequality are robust entangled states. By evenly attenuating two such light beams
by the amount $1-T$ ($T$ is the fraction of light detected), the inequality is simply transformed as
\begin{equation}
\Delta^2\hat{p}_{-,T} + \Delta^2\hat{q}_{+,T} = T(\Delta^2\hat{p}_- + \Delta^2\hat{q}_+) + 2(1-T) \; ,
\label{eq:Duanatenuado}
\end{equation}
\noindent
where $\Delta^2\hat{p}_{-,T}$, $\Delta^2\hat{q}_{+,T}$ are the new values of the EPR pair variances
after attenuation. Once violated, Eq.~(\ref{eq:DuanManco})
will remain below 2 for all values $T > 0$~\cite{PKLam2003}. In the case of uneven attenuations, a state initially violating
the inequality could in fulfill it for a finite attenuation. However, it could be brought
back to violation by attenuating the second beam, according to Eq.~(\ref{eq:Duanatenuado}). Since attenuation is a
Gaussian operation and as such cannot increase the amount of entanglement~\cite{distgauss1,distgauss2},
it must be concluded that entanglement was already present,
although not detected by the restrictive form of the Duan inequality. This demonstrates the robustness
of such entangled states. In fact, one can derive strict conditions which all entangled Gaussian states robust
against arbitrary attenuations on both fields must fulfill~\cite{ESDteorico}.

\begin{figure}[htb]
   \centering
   \includegraphics[width=4in]{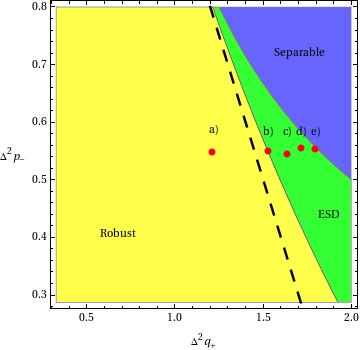}
   \caption{{\bf Space of states:} The state space is plotted as a function of the operator variances
   $\hat{p}_-$ and $\hat{q}_+$. Separable states lie in the dark
   blue region; robust entangled states are comprised within the intermediate blue region (including
   those states violating the Duan inequality); states which undergo ESD are in the lightest blue
   region. The red dots indicate the initial states produced in our experiment (Fig.~\ref{fig:ESDdata}),
   along a line of constant $\Delta^2 \hat{p}_-\approx0.50$.}
   \label{fig:ESDspace}
\end{figure}

We concentrate here on an interesting and practical situation. If the source of the entangled beams
lies with one of the parties that wish to establish secure communication, only one beam has to be sent
over a lossy channel. Then the set of robust states is enlarged, and the demands on the amount of
entanglement produced, as well as on the purity of the quantum state, are softened. In the following,
we consider a particularity of our twin beams which does not affect the essential physics of ESD but
greatly simplifies the mathematical treatment. We assume a symmetric state upon exchange of the two beams
and that no cross-quadrature (amplitude-phase) correlations exist. Then states subject to ESD fulfill the
following inequality (see Supplementary Information):
\begin{equation}
0<W_{prod}\overline{W}_{sum}+\overline{W}_{prod}W_{sum}<1\;.
\label{condESD}
\end{equation}
\noindent
Here we define $W_{sum}\equiv \Delta^2\hat{p}_- + \Delta^2\hat{q}_+ - 2$,
$\overline{W}_{sum}\equiv \Delta^2\hat{p}_+ + \Delta^2\hat{q}_- - 2$, and
$W_{prod}\equiv \Delta^2\hat{p}_-  \Delta^2\hat{q}_+ - 1$, $\overline{W}_{prod}\equiv \Delta^2\hat{p}_+  \Delta^2\hat{q}_- - 1$.
The sign of $W_{prod}$ determines whether the state is entangled ($W_{prod} < 0$) or separable ($W_{prod} \ge 0$).
The different sets of possible states are shown in Fig.~\ref{fig:ESDspace} as a function of $\Delta^2\hat{p}_-$ and $\Delta^2\hat{q}_+$,
the twin beams' squeezed variances. Entangled states subject to ESD (light blue) lie in the region
comprised between the robust states (middle-tone blue) and the separable states (dark blue).
The boundary of robust states depends on the overall purity. Two
limiting situations can be recognized: For pure states, ESD never occurs; For highly mixed states,
robustness is restricted to states violating the Duan inequality (region below the dashed line).
Since one always deals with mixed states in an experiment, it becomes important to consider the exact boundary
of Eq.~(\ref{condESD}) to assess robustness.

\begin{figure}[htb]
   \centering
   \includegraphics[width=6in]{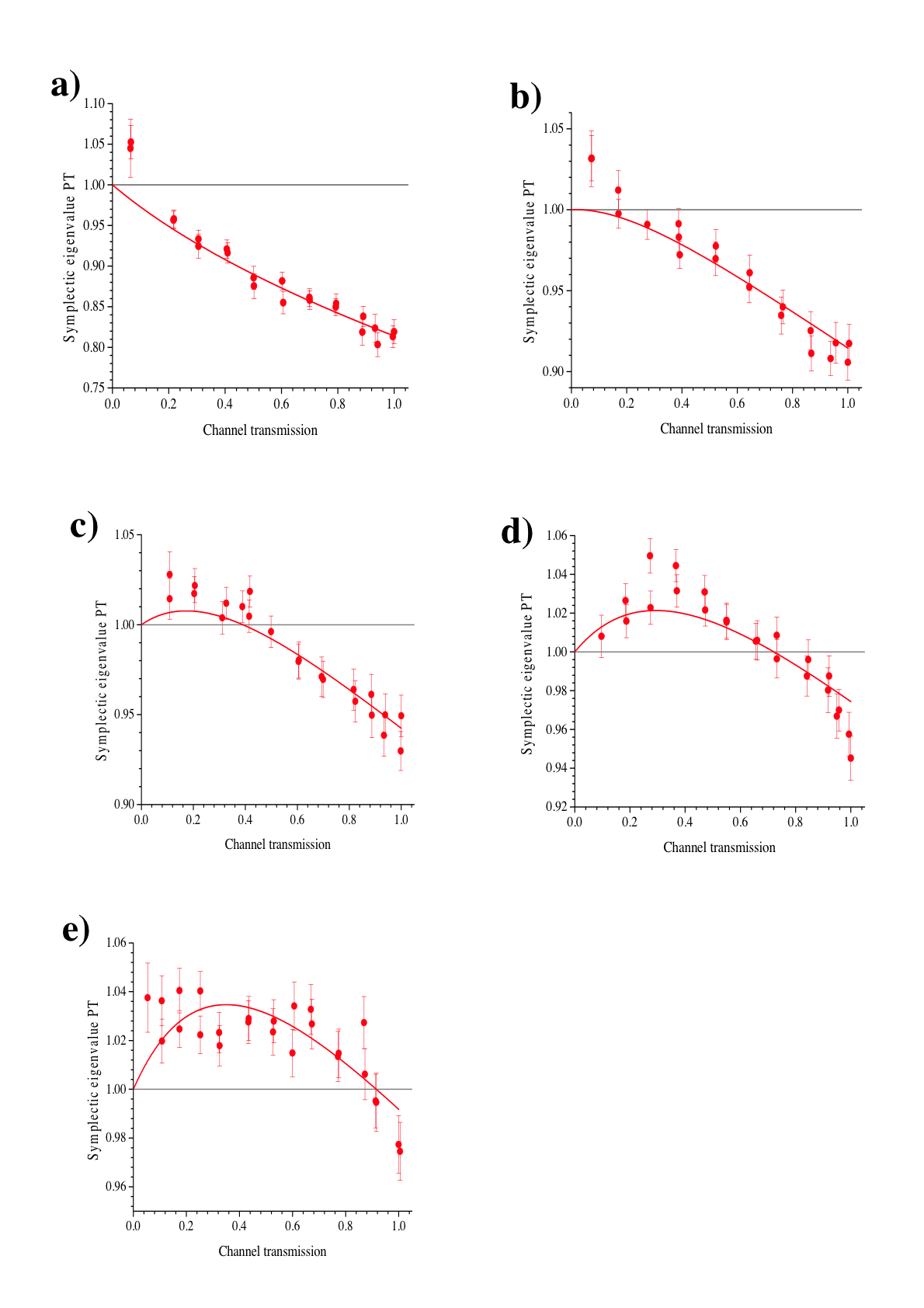}
   \caption{{\bf Entanglement data:} Symplectic eigenvalues after partial transposition as a function of the
   quantum channel transmission, for different values of pump power.
   From a) to e), as pump power increases, we observe in sequence robust entangled states, states that
   undergo ESD and, finally, an weakly entangled state. Solid lines correspond to the
   theoretical dependence of the states on losses. Good agreement is observed.}
   \label{fig:ESDdata}
\end{figure}

We generate either robust entanglement or states subject to ESD by operating the OPO under different conditions.
By varying the pump power, we control the amount of classical phonon
noise coupled to the quantum phase noise. For pump power very close to the OPO threshold,
the twin beams violate the Duan inequality and their entanglement is
robust against channel losses. As it is pumped with more power, the phase sum noise increases.
The amplitude difference noise is insensitive to the pump power ($\Delta^2 \hat{p}_-\approx0.50$).
We also checked that states experimentally produced are Gaussian, by measuring
higher than second-order moments. Details of the experiment can be found in the Supplementary
Information and references therein. The symplectic eigenvalues $\nu$ are shown in Fig.~\ref{fig:ESDdata} as
a function of the single-channel losses. The different plots correspond to several values of
the pump power, and therefore to the different initial quantum states which are indicated in Fig.~\ref{fig:ESDspace}
(red dots). The solid lines represent the expected theoretical behavior. We find good agreement.

Entanglement sudden death has attracted great attention, owing to its negative implications for
quantum information tasks. We presented here the experimental observation of ESD in a bipartite
continuous variable system, which can harm the potential applications to quantum optical communications.
In spite of the possibility of its occurrence, we could clearly draw the boundaries of those entangled states
which do not suffer ESD. This boundary can be enlarged in certain communication schemes, decreasing the
demands on the amounts of entanglement and state purity. If a ``bright future for quantum communications''
can be foreseen~\cite{PKLamNPhoton}, we have shown here how it can be made robust against losses
in the quantum channel.


\begin{thebibliography}{99}

\bibitem{QKD} Scarani, V., Bechmann-Pasquinucci, H., Cerf, N., Dusek, M., Lutkenhaus, N. \&
Peev, M. The security of practical quantum key distribution. {\em Rev. Mod. Phys.} {\bf 81}, 1301 (2008).

\bibitem{KimbleTele} Furusawa A., Sorensen, J. L., Braunstein, S. L., Fuchs, C. A., Kimble, H. J.
\& Polzik, E.S. Unconditional quantum teleportation, {\em Science} {\bf 282}, 706 (1998).

\bibitem{KimbleQuInternet} Kimble, H. J.  The quantum internet, {\em Nature} {\bf 453}, 1023 (2008).

\bibitem{PKLamNPhoton} Ralph, T. C. \& Lam, P. K., A bright future for quantum communications,
{\em Nature Photon.} {\bf 3}, 671 (2009).

\bibitem{Eberly} Yu, T. \& Eberly, J. H. Sudden death of entanglement, {\em Science} {\bf 323}, 598 (2009).

\bibitem{UFRJ} Almeida, M. P., de Melo, F., Hor-Meyll, M., Salles, A., Walborn, S. P., Ribeiro, P. H. S.
\& Davidovich, L. Environment-induced sudden death of entanglement, {\em Science} {\bf 316}, 579 (2007).

\bibitem{Science09} A. S. Coelho, F. A. S. Barbosa, K. N. Cassemiro, A. S. Villar, M. Martinelli, and
P. Nussenzveig, Three-Color Entanglement, {\em Science} {\bf 326}, 823 (2009). Published Online
September 17, 2009 (DOI:10.1126/science.1178683).

\bibitem{epr35} A. Einstein, B. Podolsky, and N. Rosen, Can Quantum-Mechanical Description of Physical
Reality Be Considered Complete?, {\em Phys. Rev.} {\bf 47}, 777 (1935).

\bibitem{schrodPCS35} E. Schr\"odinger, Discussion of Probability Relations between Separated Systems,
{\em Proc. Camb. Philos. Soc.} {\bf 31}, 555 (1935).

\bibitem{bachor} H. A. Bachor, T. C. Ralph, A Guide to Experiments in Quantum Optics (Wiley-VCH,
ed. 2, 2004).

\bibitem{ReidDrumm88} M. D. Reid and P. D. Drummond,
Quantum Correlations of Phase in Nondegenerate Parametric Oscillation,
{\em Phys. Rev. Lett.} \textbf{60}, 2731 (1988).

\bibitem{prl05} A. S. Villar, L. S. Cruz, K. N. Cassemiro, M. Martinelli, and P. Nussenzveig,
Generation of Bright Two-Color Continuous Variable Entanglement,
{\em Phys. Rev. Lett.} {\bf 95}, 243603 (2005).

\bibitem{pra09} J. E. S. César, A. S. Coelho, K. N. Cassemiro, A. S. Villar, M. Lassen, P.
Nussenzveig, M. Martinelli, Extra phase noise from thermal fluctuations in nonlinear optical crystals,
{\em Phys. Rev. A} {\bf 79}, 063816 (2009)

\bibitem{simon} R. Simon,
Peres-Horodecki Separability Criterion for Continuous Variable Systems,
{\em Phys. Rev. Lett.} \textbf{84}, 2726 (2000).

\bibitem{duan} L. M. Duan, G. Giedke, J. I. Cirac, and P. Zoller,
Inseparability criterion for continuous variable systems,
{\em Phys. Rev. Lett.} \textbf{84}, 2722  (2000).

\bibitem{PKLam2003} W. P. Bowen, R. Schnabel, P. K. Lam, and T. C. Ralph,
Experimental Investigation of Criteria for Continuous Variable Entanglement,
{\em Phys. Rev. Lett.} \textbf{90}, 043601 (2003).

\bibitem{distgauss1} J. Eisert, S. Scheel, and M. B. Plenio, Distilling Gaussian States
with Gaussian Operations is Impossible, {\em Phys. Rev. Lett.} \textbf{89}, 137903 (2002).

\bibitem{distgauss2} G. Giedke and J. I. Cirac, Characterization of Gaussian operations and
distillation of Gaussian states, {\em Phys. Rev. A} \textbf{66}, 032316 (2002).

\bibitem{ESDteorico} F. A. S. Barbosa, A. S. Coelho, A. J. de Faria, K. N. Cassemiro, A. S. Villar,
P. Nussenzveig, and M. Martinelli, {\em in preparation}.

\end{thebibliography}
\end{document}